\documentclass[11pt,a4paper]{article}
\usepackage[utf8]{inputenc}
\usepackage[left=2.5cm,top=3cm,right=2cm,bottom=3cm]{geometry}
\usepackage{authblk}
\usepackage{etex}
\usepackage[T1]{fontenc}
\usepackage{pst-pdf, pst-all, amsthm, amssymb, latexsym}
\usepackage{mathtools}
\usepackage{graphicx}
\usepackage{fancyhdr}
\usepackage{fancybox}
\usepackage{shadow}
\usepackage{empheq}
\usepackage{titlesec}
\usepackage{titletoc}
\usepackage{soul, color}
\usepackage{float}
\usepackage{shorttoc}
\usepackage{xcolor}
\usepackage{fancybox}
\usepackage{array}
\usepackage{tabularx}
\usepackage{booktabs}
\usepackage{rotating}
\usepackage{bigstrut}
\usepackage{aeguill}

\usepackage{changepage}
\newenvironment{changemargin}[2]{\begin{list}{}{%
\setlength{\topsep}{0pt}%
\setlength{\leftmargin}{0pt}%
\setlength{\rightmargin}{0pt}%
\setlength{\listparindent}{\parindent}%
\setlength{\itemindent}{\parindent}%
\setlength{\parsep}{0pt plus 1pt}%
\addtolength{\leftmargin}{#1}%
\addtolength{\rightmargin}{#2}%
}\item }{\end{list}}

\usepackage{hyperref}
\def\sym#1{\ifmmode^{#1}\else\(^{#1}\)\fi}
\usepackage{caption}
\usepackage{blindtext}
\usepackage[utf8]{inputenc}


\usepackage{caption}
\usepackage{subcaption}
\usepackage{mathtools}

\usepackage{graphicx}

\usepackage{enumerate}

\usepackage{hyperref}
\hypersetup{backref, colorlinks=true, citecolor=blue!50!black}
\usepackage{graphicx}

\usepackage{lscape} 
\renewcommand{\thetable}{\Roman{table}}
\renewcommand{\thefigure}{\Roman{figure}}

\usepackage{titlesec}
\usepackage{setspace} 
\usepackage{indentfirst}

\makeatletter
  \newcommand\smalls{\@setfontsize\smalls{10.3pt}{6}}
\makeatother

\makeatletter
  \newcommand\footnotesizes{\@setfontsize\footnotesizes{9.6pt}{6}}
\makeatother

\newtheorem*{prop1}{Proposition 1} 
\newtheorem*{prop2}{Proposition 2}

\usepackage{scalerel}
\newcommand\vfrac[2]{\ThisStyle{%
  \setbox0=\hbox{$\SavedStyle#1#2$}%
  \setbox2=\hbox{$\SavedStyle X$}%
  \ifdim\ht0>\ht2\setlength{\ht0}{\ht2}\fi%
  #1\mathord{\stretchto{\raisebox{2.3\LMpt}{$\SavedStyle/$}}{\ht0}}#2}}

\titleformat{\subsubsection}
  {\normalfont\fontsize{12}{17}\selectfont}{\thesubsubsection}{1em}{}

\usepackage[authoryear]{natbib}
\newsavebox\tmpbox

\usepackage{caption}
\usepackage{subcaption}

\begin{document}

\title{Ethnic Groups' Access to State Power and Group Size}

\author[1]{Hector Galindo-Silva\thanks{galindoh@javeriana.edu.co. I thank participants at several seminars for their helpful suggestions.  All errors or omissions are mine.}}
\affil[1]{\small Department of Economics, Pontificia Universidad Javeriana}

\maketitle
\vspace{-2pt}

\begin{abstract}
Ethnic-based political inequality is widespread, yet its underlying drivers remain poorly understood. This paper shows that an ethnic group's relative size is a key correlate of its access to central executive power. Using data on 575 groups across 181 countries from 1946 to 2021, I document a robust inverted-U-shaped relationship: groups of intermediate size are significantly more likely to gain political inclusion than both very small and very large ones. A simple model explains this pattern as the result of elite trade-offs between the risks of conflict from exclusion and the costs of sharing political rents. The model further predicts—and the data confirm—that the inverted-U is most pronounced in countries with historically competitive institutions. These findings offer new insight into the joint role of ethnic composition and institutions in shaping patterns of ethnic political inclusion.
\end{abstract}


\newpage

\section{Introduction}

Ethnic diversity characterizes most modern states, regardless of development level.\footnote{See \cite{Alesina2003DevleeschauwerEasterlyKurlatWacziarg2003} and \cite{Fearon2003}, who propose widely used measures of ethnic fractionalization and document substantial cross-country variation. \cite{PatsiurkoCampbellHall2012} provide time-varying measures showing that even historically homogeneous countries have become increasingly diverse. On the political consequences of ethnic divisions, \cite{VogtBormannRueggerCedermanHunzikerGirardin2015} introduce the Ethnic Power Relations (EPR) dataset—which I use extensively in this paper—to study how ethnic groups gain or lose access to state power.} Yet despite potential benefits from diversity—including enhanced productivity through trade and innovation \citep{AlesinaLaFerrara2005, PoschSchulzHenrich2025}—systematic political exclusion of ethnic minorities remains widespread. Between 1946 and 2021, over 35\% of politically relevant ethnic groups experienced active discrimination, and nearly 60\% were excluded from central executive power at some point.\footnote{Based on EPR Core Dataset 2021. The United Nations has long emphasized that formal participation rights are insufficient without meaningful influence over decision-making (Resolution 47/135, December 18, 1992; OHCHR Forum on Minority Issues, 2009).} Understanding which groups gain access to state power—and why—is therefore crucial for both theory and policy.

This paper documents a robust empirical regularity: the relationship between ethnic group size and access to central political power follows an inverted-U shape. Using data on 575 ethnic groups across 181 countries from 1946 to 2021, I show that groups of intermediate size are most likely to gain political inclusion, while both very small and very large groups face systematically higher exclusion rates. An ethnic group comprising approximately 40\% of a country's population has roughly 30 percentage points higher probability of accessing central power than a group comprising either 10\% or 80\% of the population.

This finding is remarkably robust. The inverted-U pattern persists across alternative temporal aggregations (five-year and ten-year periods), different outcome measures (a six-point access index and binary inclusion indicators), demanding fixed effects structures (country-by-period fixed effects), and instrumental variable strategies using lagged population shares to address simultaneity. The relationship is not driven by particular regions, time periods, or coding decisions in the underlying data. Rather, it represents a systematic global pattern in how group demographics shape political inclusion.

Why might intermediate-sized groups enjoy privileged access? I develop a simple model in which incumbent elites trade off two considerations when deciding whether to include an excluded group. On one hand, exclusion risks costly conflict, and larger excluded groups pose greater threats. On the other hand, inclusion means sharing political rents, and this cost rises with the included group's size. For very small groups, conflict threats are negligible, so exclusion is cheap. For very large groups, the numerically small incumbent elite would sacrifice substantial per capita rents by sharing power. Only intermediate-sized groups generate sufficient conflict pressure to compel inclusion without triggering prohibitive rent dilution.

The model yields a testable comparative static: institutional persistence should moderate the size-access relationship. Where political hierarchies are rigid—so that formal inclusion does not translate to proportional influence—incumbents face lower costs from granting access. The inverted-U pattern should therefore flatten in countries with persistent institutional barriers to political mobility.

I test this prediction using historical measures of executive recruitment competitiveness from Polity V, computing each country's average institutional openness since 1800 (or independence). The results strongly support the theoretical mechanism. In countries with historically competitive institutions, the inverted-U pattern is pronounced: access peaks sharply around 40\% group size and declines thereafter. In countries with historically rigid institutions, the relationship is nearly monotonic—larger groups simply have greater access, with no downturn. This heterogeneity suggests that the strategic logic of rent protection operates primarily where formal power-sharing carries real consequences.

This paper makes three contributions. First, it provides the most comprehensive evidence to date on the relationship between ethnic group size and political inclusion. While \cite{FrancoisRainerTrebbi2015} document similar nonlinearities for cabinet representation in Sub-Saharan Africa, this paper extends the finding globally, covers a longer time horizon (75 versus 40 years), and focuses on access to central executive power. The global scope matters: Sub-Saharan Africa's distinctive colonial legacies and post-independence trajectories may limit generalizability, and confirming that the pattern holds worldwide strengthens both empirical and theoretical conclusions.

Second, the paper speaks to long-standing debates about group threat versus intergroup contact. The group threat hypothesis \citep{Blumer1958, Blalock1967} predicts greater discrimination as minority populations grow, while the contact hypothesis \citep{Allport1954, Pettigrew1998} argues that larger groups facilitate prejudice-reducing interaction. Prior evidence has been mixed, partly due to identification challenges \citep{SchlueterScheepers2010}. The inverted-U finding suggests both mechanisms operate simultaneously: small groups face exclusion consistent with limited contact, while large groups face exclusion consistent with heightened threat, with intermediate groups balancing these forces.

Third, the institutional heterogeneity results offer new evidence on how historical institutions shape contemporary political outcomes. The finding that the size-access relationship differs fundamentally between institutionally open and closed societies connects to broader literatures on institutional persistence \citep{AcemogluJohnsonRobinson2001} and the political economy of ethnic inclusion.

The findings carry direct policy implications. Standard power-sharing arrangements assume demographic representation translates to political influence, but this may fail for very large or very small groups. Large minorities may require protections against majoritarian exclusion—such as supermajority requirements or guaranteed representation—while small groups may need alternative mechanisms like reserved seats or veto rights. More broadly, formal inclusion is insufficient if underlying institutions continue favoring incumbents; effective power-sharing requires complementary reforms ensuring that access translates to influence.

Several limitations warrant acknowledgment. Observational data cannot definitively establish causality, and the model abstracts from factors such as intra-elite divisions, international pressures, and group characteristics beyond size. The analysis focuses on central executive power rather than legislative representation or subnational politics. These qualifications notwithstanding, the consistency of results across diverse specifications and country contexts provides strong evidence that group size systematically shapes political access in ways that existing theories do not fully explain.

The paper proceeds as follows. Section \ref{dataandidentification} describes the data and empirical strategy. Section \ref{baselineresults} presents the main results documenting the inverted-U relationship. Section \ref{model} develops the theoretical framework. Section \ref{additionalevidence} tests the model's prediction regarding institutional persistence. Section \ref{conclusion} concludes.

 
\section{Data and Empirical Strategy}\label{dataandidentification}

\subsection{Data}

The analysis uses the Ethnic Power Relations (EPR) Core Dataset 2021 \citep{VogtBormannRueggerCedermanHunzikerGirardin2015}, which identifies politically relevant ethnic groups and documents their access to state power worldwide from 1946 to 2021.\footnote{Publicly available at \url{https://icr.ethz.ch/data/epr/core/}. \cite{BesleyMueller2018} use this dataset to study a related topic.} The EPR defines ethnicity as subjectively experienced commonality based on shared ancestry and culture, encompassing ethnolinguistic, ethnoreligious, and ethnosomatic (``racial'') groups. Coverage includes all countries with populations exceeding 250,000 at any point during the sample period.\footnote{A key advantage over alternative datasets—such as the Ethnographic Atlas or the Ancestral Characteristics database \citep{GiulianoNunn2018}—is the EPR's explicit focus on politically relevant groups with systematic documentation of power access over time. The Ancestral Characteristics database covers approximately 1,300 ethnic groups but relies on largely time-invariant population information and does not track political status.}

A group is politically relevant if at least one significant actor claims to represent it in national politics, or if its members face systematic discrimination. Political access is coded using three broad categories—monopoly/dominance, power-sharing, and exclusion—with finer subdivisions. Groups that rule alone hold either monopoly or dominant power in the executive. Groups that share central power participate as senior or junior partners in government. Excluded groups include those that are powerless (no influence but not explicitly discriminated against), actively discriminated against, self-excluded, or politically irrelevant (ethnicity not politicized despite sufficient size).

These qualitative categories imply a clear hierarchy. ``Monopoly'' represents maximum access, followed by ``dominance,'' which exceeds any power-sharing arrangement. Among excluded groups, ``powerless'' generally reflects greater access than irrelevance or self-exclusion, while discrimination represents the most severe exclusion form.\footnote{The dataset does not specify why groups become irrelevant; some may reach this status after sustained discrimination. The relative position of self-exclusion is ambiguous.}

I construct a 0–5 index mapping these categories to numeric values (Table \ref{indexaccesspower_tab}).

\begin{table}[H]
  \centering
  \caption{Access to Power Index}\label{indexaccesspower_tab}
  \vspace{-0.3cm}
  \begin{tabular}{|l|c|}
\hline
EPR Classification & Access Score \\
\hline
Monopoly & 5 \\
Dominance & 4 \\
Senior partner & 3 \\
Junior partner & 2 \\
Powerless / Irrelevance / Self-exclusion & 1 \\
Discrimination & 0 \\
\hline
\end{tabular}
\begin{minipage}{11cm}
\footnotesize \textit{Note:} Index assigns higher values to greater political access. Powerless, irrelevant, and self-excluded groups receive identical scores (1) reflecting similarly limited influence. Discrimination (0) represents complete exclusion with active state hostility.
\end{minipage}
\end{table}

This scale embodies two key assumptions. First, actively discriminated groups have less access than powerless, irrelevant, or self-excluded groups. Active discrimination creates stronger barriers than mere lack of influence. Second, powerless, irrelevant, and self-excluded groups are assigned equal scores. If a group is powerless or self-excluded, dominant groups likely view its members as having negligible prospects for gaining central political influence—functionally equivalent to irrelevance.\footnote{Section \ref{baselineresults} examines robustness to alternative measures, including binary inclusion/exclusion indicators.}

The EPR provides each group's relative population size within its country. This variable exhibits substantial variation—both cross-sectionally (from very small minorities like the San peoples in Southern Africa to clear majorities like Armenians in Armenia) and over time. Approximately 75\% of groups have multiple temporal observations, enabling not only country-by-period fixed effects specifications but also, in some specifications, group fixed effects.

I supplement the EPR with institutional data from Polity V, specifically the ``Competitiveness of Executive Recruitment'' (xrcomp) variable. This measures the extent to which institutional arrangements provide subordinates equal opportunities to become superordinates, coded 0–3 with 3 representing maximum competitiveness. To capture institutional persistence, I compute for each country-year the historical average since 1800 (or independence, if later).\footnote{Section \ref{additionalevidence} discusses this measure in detail.}

\subsection{Empirical Strategy and Identification}

The literature offers no consensus on whether larger groups are more or less likely to hold political power. I therefore estimate flexible specifications allowing for nonlinear relationships. The baseline model is:
\begin{equation}
\label{eqbaseline}
access\_to\_power_{gct}=\alpha+\beta_1 \times size_{gct} +\beta_2 \times size^2_{gct}+ \gamma_{ct}+\epsilon_{gct}
\end{equation}
where $access\_to\_power_{gct}$ is the 0–5 index for group $g$ in country $c$ during five-year period $t$, and $size_{gct}$ is the group's population share. I restrict analysis to non-dominant groups (access score $\leq$ 2), as dominant groups' size-power relationship likely differs fundamentally.\footnote{Dominant groups, by definition, already control central power. Their demographic size may reflect historical consolidation rather than the strategic inclusion dynamics that the model emphasizes.} The squared term captures potential nonlinearities, with coefficients $\beta_1$ and $\beta_2$ identifying the shape of the size-access relationship.

Country-by-period fixed effects $\gamma_{ct}$ absorb all institutional, economic, and ethnopolitical factors varying at the country-period level—including democracy levels, economic development, conflict intensity, and national ethnic fractionalization. This is a demanding specification: identification comes from comparing groups of different sizes within the same country and time period. Some specifications add group fixed effects, leveraging within-group variation in size over time, or control for the group's presence in other countries, addressing the possibility that transnational ethnic networks affect both population size and political access through migration or external support. While country-by-period fixed effects absorb country-level confounders, group-specific factors—such as internal cohesion or inherited trust \citep{AlganCahuc2010}—remain a potential concern.

Political access might also affect group size: groups facing discrimination could experience lower fertility or higher out-migration, creating reverse causality. I address this in two ways. First, I re-estimate Equation (\ref{eqbaseline}) replacing size and size-squared with their lagged values (from the previous five-year period). This captures persistent effects of group size while reducing simultaneity concerns. Second, I instrument current size with lagged size. The exclusion restriction requires that lagged size affects current political access only through current size—plausible if political decisions respond to contemporaneous demographic realities rather than historical ones.

A potential concern with the IV strategy is that historical group size may directly influence current political arrangements, violating the exclusion restriction. For example, past demographic configurations could have shaped constitutional provisions, electoral rules, or informal power-sharing norms that persist independently of current population shares. However, this concern is mitigated by the inclusion of country-by-period fixed effects, which absorb time-varying institutional factors at the country level. Moreover, the five-year lag is relatively short, limiting the scope for institutional crystallization around historical demographics. The similarity between OLS and IV estimates further suggests that any direct effect of lagged size on current access—beyond its effect through current size—is small. Nonetheless, other unobserved factors could also link past size to current political access, so these estimates should be interpreted with caution.\footnote{Results are also robust to using ten-year periods with country-by-decade fixed effects.}

Although observational data cannot definitively rule out all confounders, the consistency of results across specifications—including demanding fixed effects structures, lagged variables, and instrumental variable approaches—substantially mitigates these concerns.


\section{Main Results}\label{baselineresults}

\begin{figure}[H]
\begin{center}
\captionsetup{font={normalsize,bf}}
\caption{Scatter plot of relative size and access to central power}\label{integrVSsizeOLSav10yr_fig}
\resizebox{11cm}{8cm}{\includegraphics[width=4in]{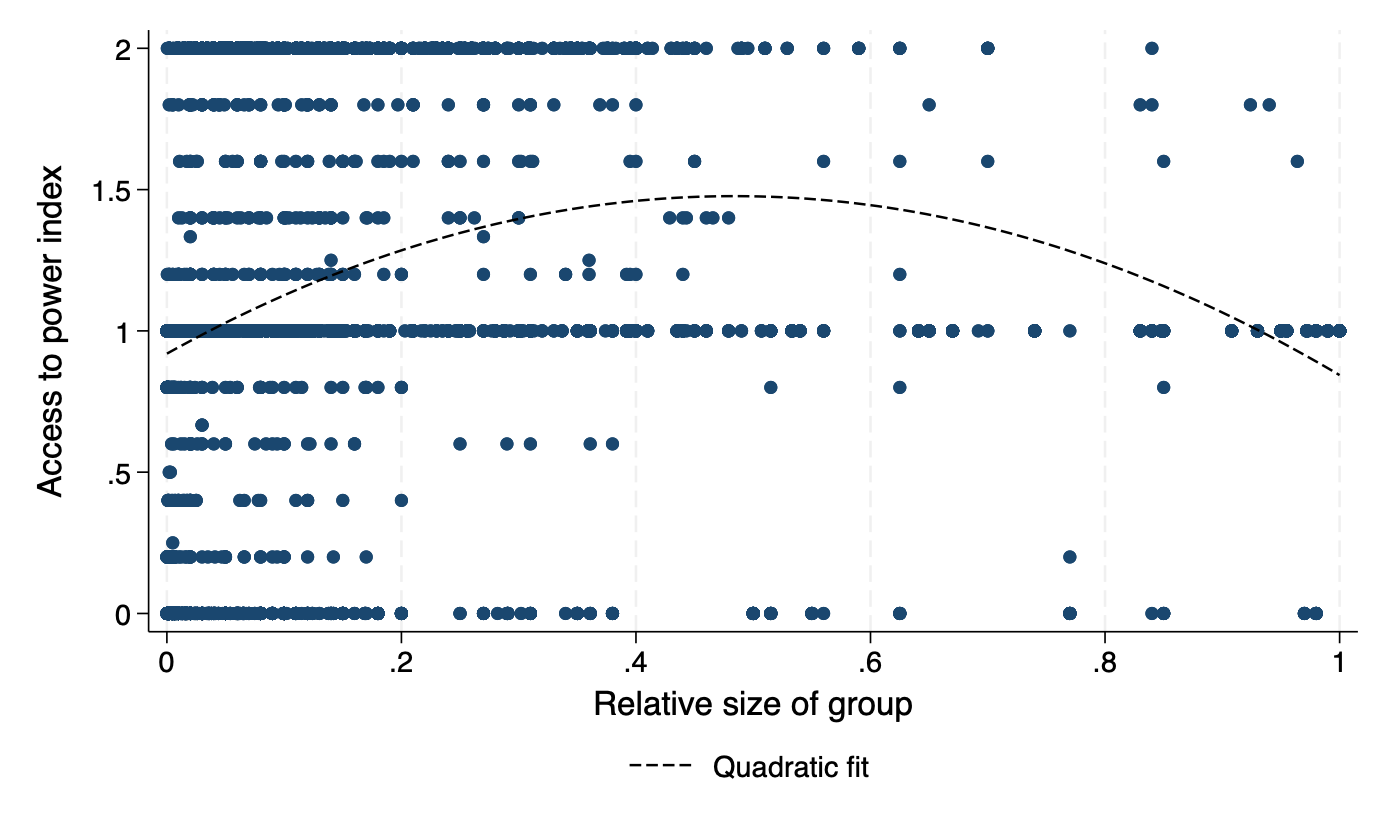}}
\begin{minipage}{11cm}
\footnotesize
\textbf{Note:} The figure plots a measure of the size of each ethnic group on the x-axis against each group's access to  power score --- as defined in Table \ref{indexaccesspower_tab} --- on the y-axis. Each point represents a group over a 5-year period. The quadratic curve that is overlaid is reported in column (1) of Table \ref{integrVSsizeOLSav10yr_tab}.
\end{minipage}
\end{center}
\end{figure}

Figure \ref{integrVSsizeOLSav10yr_fig} plots, for each non-dominant minority group in each five-year period, the relationship between the group's access-to-power score and its relative size (i.e., its share of the country's population). The figure reveals a clear inverted-U-shaped pattern.

Table \ref{integrVSsizeOLSav10yr_tab} reports estimates from variants of Equation (\ref{eqbaseline}), where the dependent variable is a 0–5 index measuring a group's access to central political power, constructed from the Ethnic Power Relations (EPR) classification. Higher values correspond to greater access to executive power, ranging from discrimination and exclusion to monopoly control. Although the access index is ordinal, the robustness of the results to a binary inclusion measure suggests that the findings do not depend on cardinality assumptions.

Column (1) presents estimates without additional controls or fixed effects and corresponds to the specification underlying Figure \ref{integrVSsizeOLSav10yr_fig}. Column (2) includes country and five-year-period fixed effects, while column (3) adds country-by-period fixed effects. Column (4) further incorporates ethnic-group fixed effects. Across all specifications, the coefficient on relative group size is positive, while the coefficient on its squared term is negative, and both are statistically significant individually and jointly. The implied turning point of the estimated relationship lies between 0.39 and 0.55, well within the feasible range of relative group size (between 0 and 1).\footnote{Specifically, the range 0.39–0.55 corresponds approximately to the 75th to 85th percentiles of the distribution of relative group size.} These results corroborate the inverted-U-shaped relationship between group size and access to power.

Table \ref{integrVSsizeOLSav10yrrobustness1_tab} examines robustness to alternative specifications. Panel A uses one observation per five-year period rather than averaging. Results closely match Table \ref{integrVSsizeOLSav10yr_tab}, with turning points between 0.37–0.49. Panel B uses ten-year periods, reducing observations but further minimizing measurement error. The inverted-U pattern persists, with turning points between 0.39–0.62. Panel C employs a binary dependent variable: 1 if the group is included (junior partner or higher), 0 if excluded. The quadratic pattern remains, confirming the relationship does not depend on treating the ordered categories as cardinal.

Table \ref{integrVSsizeOLSav10yrlagIVcontrol_tab} addresses simultaneity and omitted variable concerns. Panel A presents two strategies. Columns (1)–(2) replace current size with lagged size (five years prior), capturing persistent size effects while reducing reverse causality from current access to current population. Columns (3)–(4) use lagged size as an instrument for current size. The IV first-stage relationships are strong, and the second-stage estimates closely match baseline results. This suggests simultaneity is not driving the inverted-U pattern.

Panel B adds controls for groups' presence in other countries, addressing the possibility that transnational ethnic networks affect both size and access. The presence-in-other-countries coefficient is small and often insignificant, and including it barely affects the size coefficients. This suggests transnational networks are not major confounders.

The inverted-U-shaped relationship between group size and political access proves remarkably robust. The pattern persists across different temporal aggregations, alternative dependent variables, various fixed effects structures, simultaneity-reducing strategies, and additional controls for transnational presence. The consistency of results across these diverse specifications suggests that relative group size meaningfully shapes access to central political power.


\section{A Simple Model of Group Size and Political Inclusion}\label{model}

This section presents a simple theoretical framework explaining the inverted-U-shaped relationship between relative group size and access to central political power documented in Section \ref{baselineresults}. The goal is to convey the model's key assumptions, intuition, and results. A formal presentation, together with all definitions and proofs, is provided in the Online Appendix. The model abstracts from factors such as intra-elite divisions, international pressures, and group characteristics beyond size. While these simplifications limit the framework's scope, they allow for clear analytical predictions that can be tested empirically.

Consider a society composed of two groups. In period 1, group 1 holds monopoly political power while group 2 is excluded. In period 2, group 1 decides whether to maintain exclusion or grant group 2 access to power. If group 2 is included, political control is shared proportionally to population size, as in a system with proportional representation. Government resources are allocated according to political control and divided equally among group members. Exclusive control therefore yields higher per capita rents to the dominant group, whereas power sharing dilutes these rents.

If group 1 restricts group 2's access to power, group 2 attempts to overthrow the government, leading to costly conflict. The probability of success depends on relative group size and incumbency advantages, reflecting differences in \emph{de facto} political power \citep{AcemogluRobinson2008AER}. The destruction generated by conflict increases with the size of the excluded group, capturing the idea that larger excluded groups have greater capacity to generate instability and economic disruption.

Group 1 compares the payoff from power sharing with the expected payoff from exclusion followed by conflict. This trade-off generates a non-monotonic relationship between group size and access to power driven by three forces. First, larger excluded groups generate more destruction, making conflict costlier for group 1 conditional on winning. Second, larger excluded groups are more likely to win revolts, reducing group 1's expected payoff from exclusion. Both forces favor inclusion of large groups. Third, including group 2 means sharing resources proportionally to population. When group 1 is small (i.e., the excluded group is large), exclusive control yields very high per capita rents to group 1 members. This force favors exclusion of large groups.

For small excluded groups, conflict costs and probabilities are low—the group is weak and destruction limited. Rent dilution concerns dominate, and group 1 excludes. As the excluded group grows, conflict becomes costlier and riskier, eventually outweighing rent dilution: inclusion becomes optimal. But when the excluded group becomes very large, group 1 is numerically small, so monopoly power yields enormous per capita gains. Despite high conflict costs, group 1 prefers gambling on exclusion. This logic is consistent with existing theoretical arguments linking minority size to vulnerability \citep{Eeckhout2006, MoroNorman2004}.

\begin{prop1}
In equilibrium, access to central political power for an initially excluded group follows an inverted-U-shaped relationship with its relative size: access is lowest for very small and very large groups and highest for intermediate-sized groups.
\end{prop1}

The analysis above assumes political institutions can transition fully from exclusion to proportional inclusion. I now relax this assumption by allowing institutions to remain biased in favor of the incumbent group even when formal inclusion occurs, reflecting institutional persistence \citep{AcemogluJohnsonRobinson2001, AcemogluJohnsonRobinson2002, BanerjeeIyer2005}.

When institutions are sufficiently persistent, granting access becomes less costly for the incumbent group—particularly when the excluded group is large—because effective political control remains disproportionately concentrated. The rent dilution effect weakens, and the inverted-U flattens toward monotonicity.

\begin{prop2}
If political institutions are sufficiently persistent, the relationship between group size and access to power is monotonic: larger groups are more likely to gain access to political power.
\end{prop2}

These results generate a testable empirical prediction: the inverted-U-shaped relationship should be strongest in countries with historically open political institutions and weaker—or absent—in countries with persistent, exclusionary institutions. Section \ref{additionalevidence} empirically evaluates this prediction.\footnote{While the model proposed in this section emphasizes elite incentives—specifically, the trade-off between sharing political rents and avoiding conflict—alternative mechanisms could also generate a similar inverted-U pattern. Some are less consistent with the data: for example, if elites co-opt very large groups to neutralize threats or include small ones for symbolic reasons, we would expect greater access at the extremes, which is not observed. A more plausible alternative centers on collective action: political inclusion may depend on a group's capacity to organize and mobilize. Very small groups may coordinate effectively but lack weight; very large groups may face internal divisions or coordination problems. Intermediate-sized groups may combine sufficient scale with cohesion, enhancing their ability to demand inclusion. While this logic differs in emphasis, it is not incompatible with our framework—mobilization capacity could underlie the threat of conflict in the model. Nonetheless, we prefer the rent-sharing logic developed here, as it offers a more general foundation and yields a clear, testable empirical implication examined in the next section: the inverted-U relationship between group size and access to power should be stronger in countries with historically open political institutions, and weaker or absent in countries where institutional barriers protect incumbents.}


\section{Additional Evidence: Institutional Heterogeneity}\label{additionalevidence}

This section tests whether the relationship between ethnic group size and access to power varies systematically with the historical openness of political institutions, as predicted in Section \ref{model}. To measure institutional openness, I use Polity V's ``Competitiveness of Executive Recruitment'' (xrcomp), which captures the extent to which prevailing rules allow subordinates equal opportunity to become superordinates. The variable ranges from 0 (least competitive) to 3 (most competitive).

To operationalize institutional persistence, I compute each country's historical average of xrcomp from 1800 (or independence, if later) to the present. I then define an indicator equal to one if this average exceeds the sample median, identifying countries with historically competitive political systems.

\begin{figure}[H]
\captionsetup{font={normalsize,bf}}
\caption{Relative size and access to central power  by level of political competitiveness}\label{intVSsizeCORRav10yrsindexrcomp_fig}
\vspace{-0.5cm}
\begin{center}
\begin{subfigure}{0.45\textwidth}
\caption{High degree of competitiveness}
\includegraphics[width=3in,height=2.6in]{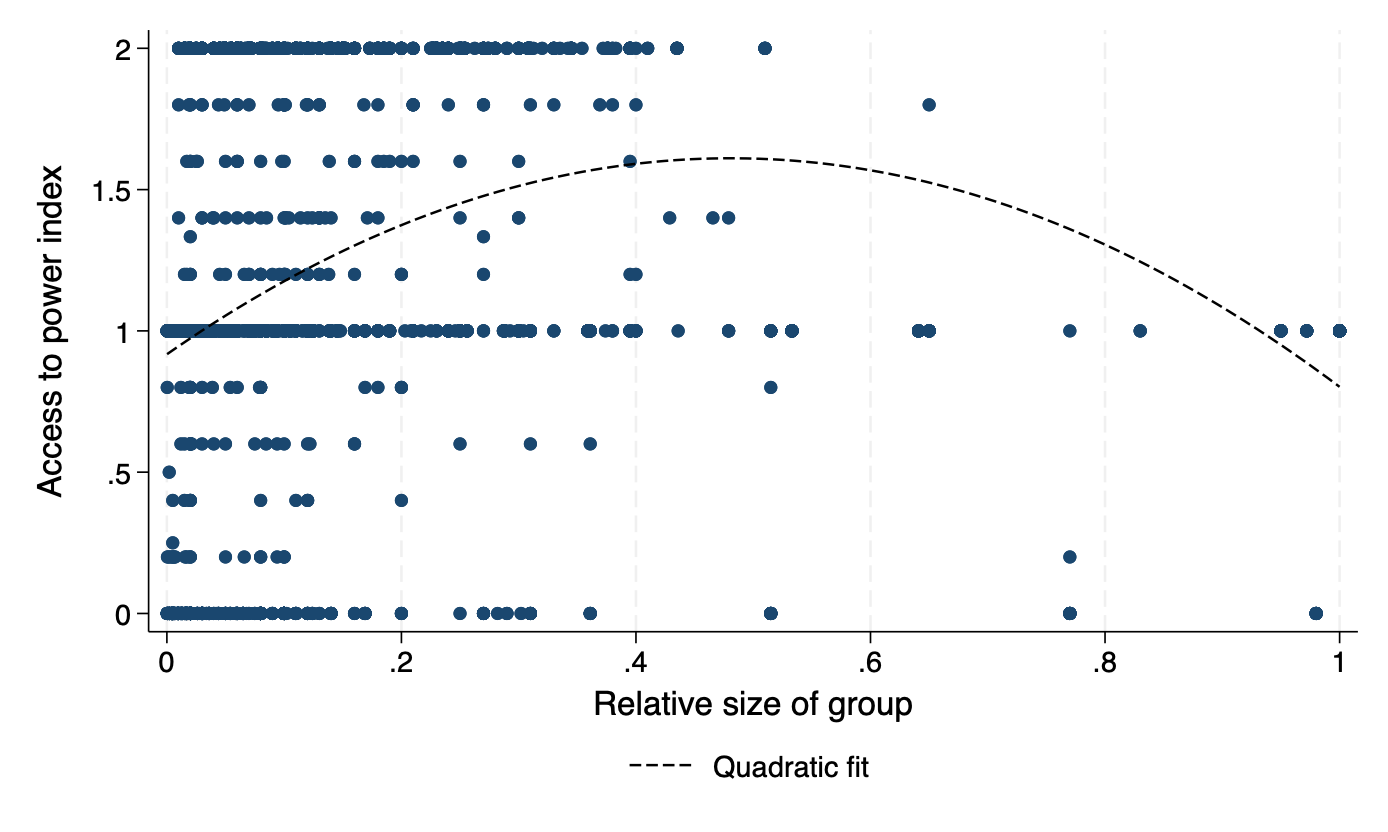}
\end{subfigure}
\begin{subfigure}{0.45\textwidth}
\caption{Low degree of competitiveness}
\includegraphics[width=3in,height=2.6in]{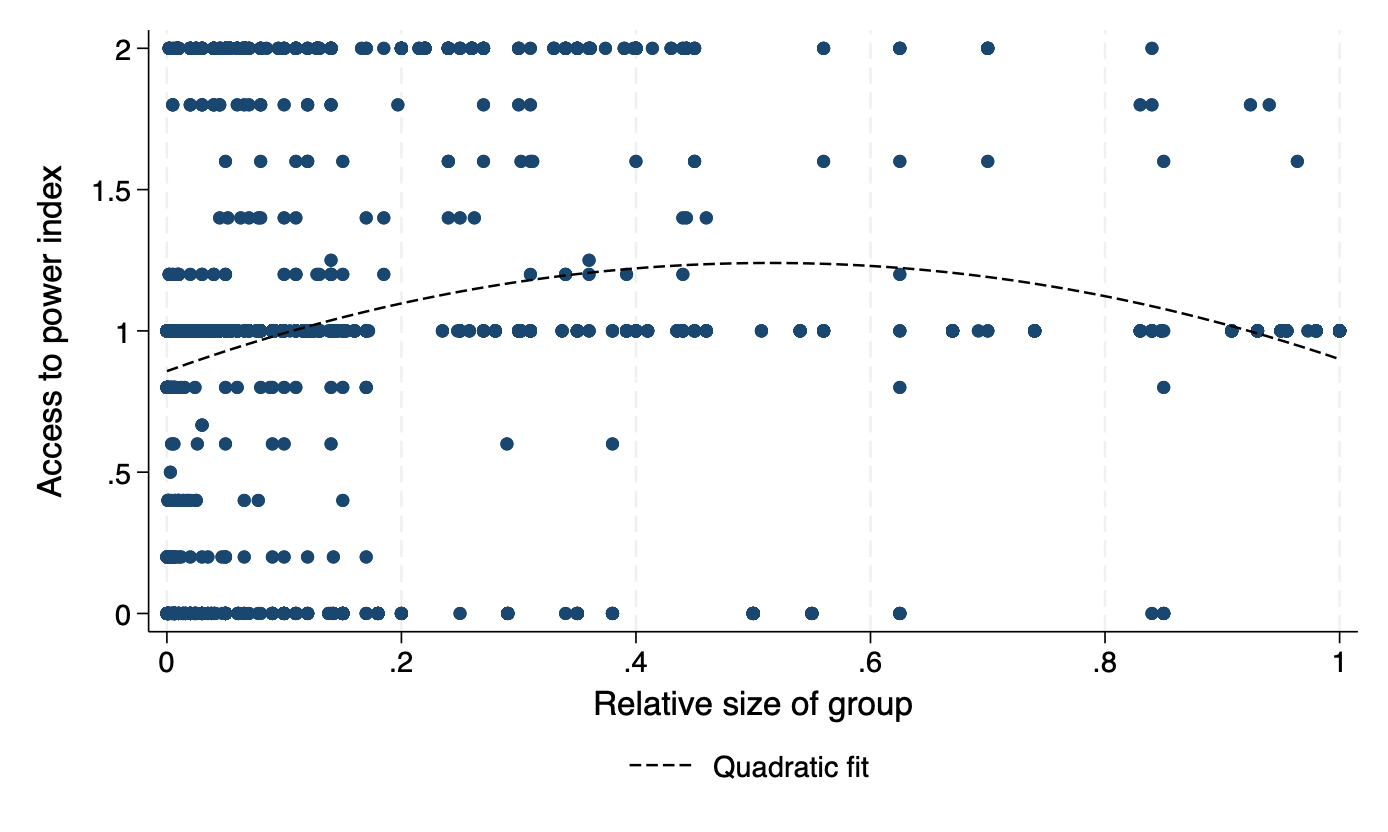}
\end{subfigure}
\begin{minipage}{15cm}  \footnotesize \textbf{Note:}  The figures plot a measure of the size of each ethnic group on the x-axis against each group's access to power score on the y-axis, by historical level of political competitiveness (i.e.\ Polity V's measure of Competitiveness of Executive Recruitment, computed since 1800).  Each point represents a group over a 10-year period. Figure (a) includes countries with an above-median level of historical political competitiveness, and Figure (b) includes countries with a below-median level of historical political competitiveness. 
\end{minipage}
\end{center}
\end{figure}

Figure \ref{intVSsizeCORRav10yrsindexrcomp_fig} plots the relationship between group size and political access separately for high- and low-competitiveness countries. Panel (a), covering historically competitive polities, shows a pronounced inverted-U shape: access peaks near 40\% group size and declines thereafter. Panel (b), covering historically less competitive polities, displays a nearly monotonic relationship: larger groups have steadily higher access.

Table \ref{integrVSsizeOLSav10yropenness_tab} formalizes this heterogeneity. Columns (1)–(2) estimate Equation (\ref{eqbaseline}) for the high-competitiveness sample, finding strong inverted-U patterns: $\beta_1 > 0$ and $\beta_2 < 0$, both statistically significant, with turning points around 0.37–0.39. Columns (3)–(4) repeat the analysis for low-competitiveness countries. Here, the quadratic term becomes small and insignificant; in some cases, it turns positive. Column (5) estimates a linear specification for this group, finding a positive and significant slope.

These patterns provide strong support for the theoretical mechanism. In countries with historically open institutions—where formal inclusion tends to translate into actual influence—incumbent elites face the rent dilution trade-off emphasized in the model, leading to the inverted-U. In contrast, where institutional barriers shield incumbents from meaningful power-sharing, the costs of including large groups are reduced, and the relationship between group size and access becomes more monotonic.


\section{Conclusion}\label{conclusion}

This paper documents a robust inverted-U-shaped relationship between ethnic group size and access to central political power. Using data on 575 groups across 181 countries from 1946 to 2021, I show that groups of intermediate size are most likely to gain political inclusion, while very small and very large groups face higher exclusion rates. This pattern persists across numerous specifications, suggesting relative group size is a key correlate and plausibly causal determinant political inequality.

A simple model attributes this nonlinearity to elite strategic calculations. Small excluded groups pose minimal conflict threats, making exclusion attractive. Large excluded groups threaten the numerically small incumbent elite's rent concentration, again favoring exclusion. Only intermediate-sized groups optimally balance conflict threats against rent dilution. The model predicts this inverted-U pattern should weaken where institutions persistently favor incumbents. Consistent with this prediction, the nonlinear relationship is pronounced in historically open polities, while countries with rigid institutional hierarchies exhibit near-monotonic patterns.

The findings carry policy implications. Standard power-sharing arrangements assume demographic representation translates to political influence, which may fail for very large or very small groups absent complementary reforms. Large minorities may require protections against majoritarian exclusion, while small groups may need alternative mechanisms like reserved seats. More broadly, formal power-sharing is insufficient if underlying institutions continue favoring incumbents; effective inclusion requires reforms ensuring formal access translates to real influence.

Several limitations warrant acknowledgment. Establishing definitive causality from observational data remains challenging. The model abstracts from intra-elite divisions, international pressures, and group characteristics beyond size. The analysis focuses on central executive power rather than legislative representation or subnational politics. Future research could examine de facto political participation, test whether other institutional features moderate the size-access relationship, and track how the relationship evolves following democratic transitions.

\hbox {}
\hbox {} \newpage
\section*{Graphs and Tables}


\begin{table}[H]
{ 
\renewcommand{\arraystretch}{0.8} 
\setlength{\tabcolsep}{10pt}
\captionsetup{font={normalsize,bf}}
\caption {Relative size and access to central power}  \label{integrVSsizeOLSav10yr_tab}
\vspace{-0.5cm}
\begin{center}  
\begin{tabular}{lcccc}
\hline\hline  \addlinespace[0.15cm]
    & \multicolumn{4}{c}{Dep. variable: Level of access to central power} \\\cmidrule[0.2pt](l){2-5}
& (1)& (2) & (3)& (4) \\  \addlinespace[0.15cm] \hline \addlinespace[0.15cm] 
Relative size       &       2.304\sym{***}&       2.927\sym{***}&       3.196\sym{***}&       2.813\sym{***}\\
                    &     (0.336)         &     (0.405)         &     (0.441)         &     (0.594)         \\
Relative size squared&      -2.379\sym{***}&      -3.489\sym{***}&      -4.131\sym{***}&      -2.545\sym{**} \\
                    &     (0.329)         &     (0.527)         &     (0.725)         &     (1.092)         \\
Joint significance  &       0.000         &       0.000         &       0.000         &       0.000         \\
Vertex              &       0.484         &       0.420         &       0.387         &       0.553         \\
R-squared           &       0.066         &       0.463         &       0.607         &       0.856         \\
Observations        &        8300         &        8298         &        7466         &        7458         \\

\addlinespace[0.15cm]\hline\addlinespace[0.15cm]           
  Country and period fixed effects & N & Y  & N & Y \\      
  Country $\times$ period fixed effects & N & N & Y & Y \\       
 Group fixed effects & N & N & N & Y  \\    
      \addlinespace[0.15cm] \hline\hline         
\multicolumn{5}{p{13.4cm}}{\footnotesize{\textbf{Notes:} 
All columns report OLS estimates for estimates from Eq (\ref{eqbaseline}). The dependent variable in all columns is each ethnic group's access to power score (as defined in Table \ref{indexaccesspower_tab}). The sample is limited to groups with a score of 2 or less, and the index is averaged over a 5-year period.  In this subsample, the average relative size is 0.117 (with s.d. 0.224) and the average level of access to central power is 1.036 (with s.d. 0.575).  Robust standard errors  clustered by country are in parentheses. * denotes results are statistically significant at the 10\% level, ** at the 5\% level, and *** at the 1\% level.} } \\
\end{tabular}
\end{center}
}
\end{table}


\begin{table}[H]
{ 
\renewcommand{\arraystretch}{0.8} 
\setlength{\tabcolsep}{10pt}
\captionsetup{font={normalsize,bf}}
\caption {Relative size and access to central power (robustness to alternatives measures of access to central power)}    \label{integrVSsizeOLSav10yrrobustness1_tab}
\begin{center}
\vspace{-0.6cm} \begin{tabular}{lccccc}
\hline\hline  \addlinespace[0.15cm]
    & \multicolumn{4}{c}{Dep. Var.: Access to central power} \\\cmidrule[0.2pt](l){2-5}
& (1)& (2) & (3)& (4) \\    \addlinespace[0.15cm] \hline \addlinespace[0.15cm] 
\emph{\underline{Panel A}:} &\multicolumn{4}{c}{First ob. within each 5-year period}     \\\cmidrule[0.2pt](l){2-5}   \addlinespace[0.15cm]
Relative size       &       2.342\sym{***}&       2.962\sym{***}&       3.299\sym{***}&       3.831\sym{***}\\
                    &     (0.336)         &     (0.409)         &     (0.468)         &     (0.533)         \\
Relative size squared&      -2.417\sym{***}&      -3.569\sym{***}&      -4.453\sym{***}&      -4.911\sym{***}\\
                    &     (0.329)         &     (0.531)         &     (0.823)         &     (0.856)         \\
Joint significance  &       0.000         &       0.000         &       0.000         &       0.000         \\
Vertex              &       0.485         &       0.415         &       0.370         &       0.390         \\
R-squared           &       0.064         &       0.446         &       0.608         &       0.847         \\
Observations        &        7901         &        7900         &        7087         &        7075         \\
 
\addlinespace[0.15cm]\hline\addlinespace[0.15cm]       
\emph{\underline{Panel B}:}  &\multicolumn{4}{c}{Averages over 10-year periods}     \\\cmidrule[0.2pt](l){2-5}   \addlinespace[0.15cm]
Relative size       &       2.319\sym{***}&       2.858\sym{***}&       3.189\sym{***}&       2.582\sym{***}\\
                    &     (0.337)         &     (0.401)         &     (0.433)         &     (0.560)         \\
Relative size squared&      -2.382\sym{***}&      -3.313\sym{***}&      -4.093\sym{***}&      -2.098\sym{**} \\
                    &     (0.331)         &     (0.527)         &     (0.709)         &     (0.846)         \\
Joint significance  &       0.000         &       0.000         &       0.000         &       0.000         \\
Vertex              &       0.487         &       0.431         &       0.390         &       0.615         \\
R-squared           &       0.069         &       0.469         &       0.602         &       0.857         \\
Observations        &        4706         &        4704         &        4245         &        4229         \\
 
\addlinespace[0.15cm]\hline\addlinespace[0.15cm]       
\emph{\underline{Panel C}:}  &\multicolumn{4}{c}{Prob. of access to power}     \\\cmidrule[0.2pt](l){2-5}   \addlinespace[0.15cm]
Relative size       &       2.212\sym{***}&       2.315\sym{***}&       2.509\sym{***}&       2.882\sym{***}\\
                    &     (0.289)         &     (0.305)         &     (0.370)         &     (0.476)         \\
Relative size squared&      -2.441\sym{***}&      -2.796\sym{***}&      -3.190\sym{***}&      -2.974\sym{***}\\
                    &     (0.281)         &     (0.388)         &     (0.584)         &     (0.860)         \\
Joint significance  &       0.000         &       0.000         &       0.000         &       0.000         \\
Vertex              &       0.453         &       0.414         &       0.393         &       0.484         \\
R-squared           &       0.135         &       0.510         &       0.623         &       0.857         \\
Observations        &        8300         &        8298         &        7466         &        7458         \\
 
\addlinespace[0.15cm]\hline\addlinespace[0.15cm]             
  Country and period fixed effects & N & Y  & N & Y \\      
  Country $\times$ period fixed effects & N & N & Y & Y \\       
 Group fixed effects & N & N & N & Y  \\    
       \addlinespace[0.15cm] \hline\hline                       
\multicolumn{5}{p{14.1cm}}{\footnotesize{\textbf{Notes:} 
All columns report OLS estimates for estimates from Eq (\ref{eqbaseline}). The dependent variable in all columns is based on ethnic group's access to power score (as defined in Table \ref{indexaccesspower_tab}). The sample is limited to groups with a score of 2 or less. Panel A uses 10-year panel, but rather than averaging the 10-year data, it takes one observation within each sub-period (e.g one every tenth year). In this sample, the average relative size is 0.121 (with s.d. 0.228) and the average level of access to power is 1.023 (with s.d. 0.609). Panel B uses a 5-year panel. In this sample, the average relative size is 0.118 (with s.d. 0.226) and the average level of access to power is 1.030 (with s.d. 0.586). Panel C uses 10-year panel and a dichotomous measure of access to power which is equal to one if the group is not excluded from central power and zero if it is excluded. In this sample, the average probability of not being excluded from central power is 0.216 (with s.d. 0.411). Robust standard errors  clustered by country are in parentheses. * denotes results are statistically significant at the 10\% level, ** at the 5\% level, and *** at the 1\% level.} } \\
\end{tabular}
\end{center}
}
\end{table}


\begin{table}[H]
{ 
\renewcommand{\arraystretch}{0.8} 
\setlength{\tabcolsep}{10pt}
\captionsetup{font={normalsize,bf}}
\caption {Relative size and access to central power (robustness to the use lagged size, IV and control for presence of each group in other countries)}  \label{integrVSsizeOLSav10yrlagIVcontrol_tab}
\vspace{-0.5cm}
\begin{center}  
\begin{tabular}{lcccc}
\hline\hline  \addlinespace[0.15cm]
    & \multicolumn{4}{c}{Dep. Var.: Level of access to central power} \\\cmidrule[0.2pt](l){2-5}
    & (1)& (2) & (3)& (4)\\  \addlinespace[0.15cm] \hline \addlinespace[0.15cm] 
   \emph{\underline{Panel A}:}       & \multicolumn{2}{c}{Lagged effect} &\multicolumn{2}{c}{IV} \\\cmidrule[0.2pt](l){2-3}\cmidrule[0.2pt](l){4-5}
Relative size       &       3.211\sym{***}&       1.937\sym{***}&       2.281\sym{***}&       2.945\sym{***}\\
                    &     (0.459)         &     (0.453)         &     (0.348)         &     (0.448)         \\
Relative size squared&      -4.204\sym{***}&      -2.226\sym{***}&      -2.361\sym{***}&      -3.570\sym{***}\\
                    &     (0.777)         &     (0.728)         &     (0.339)         &     (0.581)         \\
Joint significance  &       0.000         &       0.000         &       0.000         &       0.000         \\
Vertex              &       0.382         &       0.435         &       0.483         &       0.412         \\
R-squared           &       0.605         &       0.856         &       0.069         &       0.467         \\
Observations        &        7266         &        7258         &        8066         &        8066         \\
 
\cmidrule[0.2pt](l){2-5} 
  Country and period fixed effects & Y & Y  & Y & Y \\       
  Country $\times$ period fixed effects & Y & Y  & N & Y \\       
 Group fixed effects & N & Y  & N & N \\    
\addlinespace[0.15cm]\hline\addlinespace[0.15cm]       
  \emph{\underline{Panel B}:}         & \multicolumn{2}{c}{Baseline specification} & \multicolumn{1}{c}{Lagged} & \multicolumn{1}{c}{IV} \\\cmidrule[0.2pt](l){2-3}\cmidrule[0.2pt](l){5-5}\cmidrule[0.2pt](l){4-4}
 Relative size       &       3.185\sym{***}&       2.798\sym{***}&       3.210\sym{***}&       2.924\sym{***}\\
                    &     (0.446)         &     (0.593)         &     (0.467)         &     (0.452)         \\
Relative size squared&      -4.122\sym{***}&      -2.516\sym{**} &      -4.201\sym{***}&      -3.525\sym{***}\\
                    &     (0.738)         &     (1.093)         &     (0.798)         &     (0.587)         \\
Countries with group&      -0.019\sym{**} &      -0.010         &      -0.019\sym{**} &      -0.017\sym{**} \\
                    &     (0.009)         &     (0.010)         &     (0.009)         &     (0.008)         \\
Joint significance  &       0.000         &       0.000         &       0.000         &       0.000         \\
Vertex              &       0.386         &       0.556         &       0.382         &       0.415         \\
R-squared           &       0.613         &       0.856         &       0.611         &       0.471         \\
Observations        &        7466         &        7458         &        7232         &        8027         \\
 
 \cmidrule[0.2pt](l){2-5}        
  Country $\times$ period fixed effects & Y & Y & Y & Y  \\       
 Group fixed effects & N & Y & N  & N  \\        
      \addlinespace[0.15cm] \hline\hline         
\multicolumn{5}{p{14cm}}{\footnotesize{\textbf{Notes:} 
All columns report OLS estimates for estimates from Eq (\ref{eqbaseline}).  The dependent variable in all columns is each ethnic group's access to power score (as defined in Table \ref{indexaccesspower_tab}). The sample is limited to groups with a score of 2 or less, and the index is averaged over a 5-year period. In this subsample, the average relative size is 0.117 (with s.d. 0.224) and the average level of access to central power is 1.036 (with s.d. 0.575).  Robust standard errors  clustered by country are in parentheses. * denotes results are statistically significant at the 10\% level, ** at the 5\% level, and *** at the 1\% level.} } \\
\end{tabular}
\end{center}
}
\end{table}


\begin{table}[H]
{ 
\renewcommand{\arraystretch}{0.8} 
\setlength{\tabcolsep}{10pt}
\captionsetup{font={normalsize,bf}}
\caption {Relative size and access to central power by level of  political competitiveness}    \label{integrVSsizeOLSav10yropenness_tab}
\vspace{-0.5cm}
\begin{center}
 \begin{tabular}{lccccc}
\hline\hline \addlinespace[0.15cm]
    & \multicolumn{4}{c}{Dep. Var.: Level of access to central power} \\\cmidrule[0.2pt](l){2-6}
& (1)& (2) & (3)& (4) & (5)\\    \addlinespace[0.15cm] \hline \addlinespace[0.15cm]     
             & \multicolumn{2}{c}{High degree of} & \multicolumn{3}{c}{Low degree of} \\\
        & \multicolumn{2}{c}{competitiveness} & \multicolumn{3}{c}{competitiveness} \\\cmidrule[0.2pt](l){2-3}\cmidrule[0.2pt](l){4-6}
 Relative size       &       3.871\sym{***}&       3.504\sym{***}&       1.874\sym{**} &       0.464         &       0.749\sym{*}  \\
                    &     (0.568)         &     (0.649)         &     (0.719)         &     (1.482)         &     (0.379)         \\
Relative size squared&      -5.207\sym{***}&      -4.546\sym{***}&      -2.064\sym{*}  &       1.244         &                     \\
                    &     (0.818)         &     (1.095)         &     (1.166)         &     (1.860)         &                     \\
Joint significance  &       0.000         &       0.000         &       0.030         &       0.175         &                     \\
Vertex              &       0.372         &       0.385         &       0.454         &      -0.186         &                     \\
R-squared           &       0.589         &       0.864         &       0.642         &       0.900         &       0.636         \\
Observations        &        4803         &        4671         &        2662         &        2652         &        2662         \\

 \addlinespace[0.15cm]
 \hline \addlinespace[0.15cm]      
  Country $\times$ period fixed effects & Y & Y & Y & Y & Y\\         
 Group fixed effects & N & Y & N & Y  & Y\\     
       \addlinespace[0.15cm] \hline\hline                       
\multicolumn{6}{p{15.3cm}}{\footnotesize{\textbf{Notes:} 
All columns report OLS estimates for estimates from Eq (\ref{eqbaseline}).  The dependent variable in all columns is each ethnic group's access to power score (as defined in Table \ref{indexaccesspower_tab}). The sample is limited to groups with a score of 2 or less, and the index is averaged over a 10-year period. In this subsample, the average relative size is 0.117 (with s.d. 0.224) and the average level of access to central power is 1.036 (with s.d. 0.575).  Columns (1) and (2) include countries with an above-median level of historical political competitiveness, and Columns (3) to (5) include countries with a below-median level of historical political competitiveness (i.e.\ Polity V's measure of Competitiveness of Executive Recruitment, computed since 1800). Robust standard errors clustered by country are in parentheses. $\dagger$ denotes results are statistically significant at the 15\% level, * at the 10\% level, ** at the 5\% level, and *** at the 1\% level.} } \\
\end{tabular}
\end{center}
}
\end{table}


\hbox {}
\hbox {} \newpage
\section*{Online Appendix: Model, Definitions, and Proofs}

\renewcommand{\thefigure}{\Alph{section}\arabic{figure}}
\renewcommand{\thetable}{\Alph{section}\arabic{table}}

This Online Appendix presents the full formal version of the model summarized in
Section~\ref{model}. The main text focuses on intuition and results, while this
appendix defines all objects precisely and provides the complete proofs of
Propositions~1 and~2. All notation, assumptions, and equations follow the main
text exactly.

\section*{A. Model environment}

Consider a society composed of individuals who belong to $n$ internally homogeneous
groups. Let $N$ be the total number of individuals in the society, with $N_i$
representing the number of individuals in group $i$, so that
\[
\sum_{i=1}^n N_i = N.
\]

There are two time periods. In period~1, one group (or a coalition of groups)
fully controls the government and decides whether to continue limiting other
groups' access to power in period~2.

Without loss of generality, set $n=2$. Group~1 has an \emph{ex ante} monopoly of
political power in period~1. Group~2 represents a group (or cluster of groups)
with limited access to power in period~1 and is \emph{de jure} excluded from
government decisions.

If group~2's access to power in period~2 is not restricted, both groups
participate in the period-2 government. Let $s_i$ denote group $i$'s level of
control over the period-2 government, defined as
\begin{equation}
\label{si}
s_i=\frac{N_i}{\sum_{j=1}^2 N_j}.
\end{equation}

\section*{B. Government payoffs under inclusion}

The government has an exogenously determined budget. First, each group that
participates in government receives an allocation of the budget proportional
to its level of political control. Second, this allocation is divided equally
among the group's members.

If group~$i$ has monopoly control of the government and the per-capita value
of the budget is normalized to unity, each member of group~$i$ receives
$N/N_i$.

Individuals use their government transfer to purchase a private good in period~2,
from which they derive utility that increases linearly in consumption.

The per-capita payoff to an individual in group~1 when group~2's access to power
is not restricted is therefore
\begin{equation}
\label{payofflimited}
s_1\left(\frac{N}{N_1}\right).
\end{equation}

\section*{C. Exclusion, conflict, and payoffs}

If group~1 restricts group~2's access to power in period~2, group~2 attempts to
overthrow the government. This results in a costly contest. The group that wins
fully controls the government in period~2.

The probability that group~$i$ wins the contest is given by
\begin{equation}
\label{pi}
p_i=\frac{N_i+a_i}{\sum_{j=1}^2 (N_j+a_j)},
\end{equation}
where $a_i\geq 0$ captures the group's ability to solve collective-action
problems and thus its \emph{de facto} power.

Conflict destroys a fraction of the government budget. This fraction is assumed
to be proportional to the relative size of group~2. Let $\lambda\in(0,1)$ denote
the marginal increase in destruction associated with group~2's size.

The per-capita payoff to an individual in group~1 when group~2's access to power
is restricted is
\begin{equation}
\label{payoffnotlimited}
p_1\left(\frac{N}{N_1}\right)\big(1-\lambda (N_2/N)\big).
\end{equation}

\section*{D. Assumptions}

Two assumptions are imposed.

First, group~1 has greater \emph{de facto} power than group~2:
\[
a_1>a_2, \qquad a_2=0.
\]

Second, conflict is sufficiently costly:
\begin{equation}
\label{assumption1}
\frac{\lambda}{1-\lambda}>a_1.
\end{equation}

This assumption guarantees that restricting group~2's access to power is not
always a dominant strategy for group~1.

\section*{E. Equilibrium condition}

Group~1 grants group~2 access to power in period~2 whenever
\begin{equation}
\label{condnotlimiting}
p_1\left(\frac{N}{N_1}\right)\big(1-\lambda (N_2/N)\big)
\geq
s_1\left(\frac{N}{N_1}\right).
\end{equation}

Normalizing $N=1$ and defining $\delta=N_2$ (so $N_1=1-\delta$),
condition~\eqref{condnotlimiting} can be rewritten as
\begin{equation}
\label{maincond}
\frac{(1-\delta+a_1)}{(1+a_1)}
\frac{(1-\lambda\delta)}{(1-\delta)}-1\geq 0.
\end{equation}

\section*{F. Proof of Proposition 1}

\begin{proof}
Define
\begin{equation}
\label{functionfdelta}
f(\delta)=
1-
\frac{(1-\delta+a_1)}{(1+a_1)}
\frac{(1-\lambda\delta)}{(1-\delta)}.
\end{equation}

Condition~\eqref{maincond} holds if and only if $f(\delta)>0$.
Differentiating,
\[
f'(\delta)=
-\frac{1}{(1+a_1)}
\frac{a_1-\lambda(1+a_1)+2\lambda\delta-\lambda\delta^2}{(1-\delta)^2},
\]
and
\[
f''(\delta)=
-\frac{2a_1(1-\lambda)}{(1+a_1)(1-\delta)^3}<0.
\]

Thus $f(\delta)$ is strictly concave on $(0,1)$ and has a unique maximum
$\delta^*\in(0,1)$ under Assumption~\eqref{assumption1}. Moreover,
$f(0)=0$ and $f(\delta)<0$ for $\delta$ sufficiently close to~1.
Therefore, access to power follows an inverted-U pattern.
\end{proof}

\section*{G. Institutional persistence}

Suppose that political institutions remain biased in favor of group~1 even when
group~2 is formally included. Let group~2's effective participation be
\begin{equation}
\label{q2}
q_2=\gamma s_2,
\end{equation}
where $\gamma\in[0,1]$ measures institutional openness.

Replacing $s_2$ with $q_2$ in~\eqref{maincond} yields
\begin{equation}
\label{democcond}
\frac{(1-\delta+a_1)}{(1+a_1)}
\frac{(1-\lambda\delta)}{(1-\delta)}
-
\frac{(1-\gamma\delta)}{(1-\delta)}
\geq 0.
\end{equation}

\section*{H. Proof of Proposition 2}

\begin{proof}
Define
\begin{equation}
\label{functionhdelta}
h(\delta)=
\gamma+\frac{(1-\gamma)}{(1-\delta)}
-
\frac{(1-\delta+a_1)}{(1+a_1)}
\frac{(1-\lambda\delta)}{(1-\delta)}.
\end{equation}

Condition~\eqref{democcond} holds if and only if $h(\delta)>0$.
Differentiating twice,
\[
h''(\delta)=
\frac{2}{(1-\delta)^3}
\left[(1-\gamma)-\frac{a_1(1-\lambda)}{(1+a_1)}\right].
\]

Define
\[
\gamma^*=\frac{1+\lambda a_1}{1+a_1}.
\]

For $\gamma<\gamma^*$, $h(\delta)$ is strictly increasing on $(0,1)$, implying
that larger groups are more likely to obtain access to power.
\end{proof}

\hbox {} \newpage

\bibliographystyle{aer}
\bibliography{ACCESSTOPOWERbibALL}

\end{document}